\documentclass[fp,twocolumn]{jpsj3}
\usepackage[dvipdfmx]{graphicx}
\usepackage{amsmath}
\usepackage{bm}
\usepackage{color}
\newcommand{\kf}{k_{\mathrm{F}}}

\newcommand{\as}{a_s}

\newcommand{\bp}{\bm{p}}
\newcommand{\bq}{\bm{q}}
\begin{document}
\title{Kovtun-Son-Starinets Conjecture and Effects of Mass Imbalance in the Normal State of an Ultracold Fermi Gas in the BCS-BEC Crossover Region}
\author{Daichi Kagamihara and Yoji Ohashi}
\inst{Department of Physics, Keio University, 3-14-1 Hiyoshi, Kohoku-ku, Yokohama 223-8522, Japan}
\date{\today}
\abst{We theoretically assess the conjecture proposed by Kovtun, Son, and Starinets, stating that the ratio $\eta/s$ of the shear viscosity $\eta$ to the entropy density $s$ has the lower bound as $\eta/s\ge\hbar/(4\pi k_{\mathrm{B}})$. In the normal state of a mass-imbalanced ultracold Fermi gas, consistently including strong-coupling corrections to both $\eta$ and $s$ within the self-consistent $T$-matrix approximation, we evaluate $\eta/s$ over the entire BCS (Bardeen-Cooper-Schrieffer)-BEC (Bose-Einstein condensation) crossover region, in the presence of mass imbalance. We find that $\eta/s$ achieves the minimum value $4.5\times \hbar/(4\pi k_{\mathrm{B}})$, not at the unitarity, but slightly in the BEC regime, $(k_{\mathrm{F}}a_s)^{-1}\simeq 0.4>0$ (where $a_s$ is the $s$-wave scattering length, and $k_{\mathrm{F}}$ is the Fermi momentum). In contract to the previous expectation, we find that this lower bound is almost independent of mass imbalance: Our results predict that all the mass-balanced $^6$Li-$^6$Li and $^{40}$K-$^{40}$K mixtures and the mass-imbalanced $^{40}$K-$^{161}$Dy mixture give almost the same lower bound of $\eta/s$. We also point out that the two quantum phenomena, Pauli blocking and bound-state formation, are crucial keys for the lower bound of $\eta/s$.}
\maketitle
\par
\par
\section{Introduction}
\par
In 2005, Kovtun, Son, and Starinets (KSS) proposed\cite{Kovtun:2005aa} that the ratio $\eta/s$ of the shear viscosity $\eta$ to the entropy density $s$ should be lower-bounded as
\begin{equation}
\frac{\eta}{s} \geq \frac{\hbar}{4\pi k_{\mathrm{B}}},
\label{eq.1}
\end{equation}
in all relativistic quantum field theories at finite temperature with zero chemical potential. Because Eq. (\ref{eq.1}) doesn't involve the speed of light $c$, KSS predicted that it would also be applicable to the non-relativistic case, at least in a single-component gas with spin 0 or 1/2.
\par
Regarding this so-called KSS conjecture, the factor $\hbar$ in the right-hand side of Eq. (\ref{eq.1}) indicates that the existence of the lower bound (KSS bound) is associated with a quantum effect. In addition, we note that this bound is also related to particle-particle correlations. To simply see this, we conveniently employ the expressions for the viscosity, 
\begin{equation}
\eta\sim nl_{\mathrm{mfp}}p_{\mathrm{av}},
\label{eq_simple}
\end{equation}
as well as the entropy density,
\begin{equation}
s\sim nk_{\mathrm{B}},
\end{equation}
in a simple classical gas\cite{Bruun:2005aa,Massignan:2005aa}. Here, $n$, $l_{\mathrm{mfp}}$, and $p_{\mathrm{av}}$ are the number density, the mean free path, and the averaged momentum, respectively. Then, the ratio,
\begin{equation}
\frac{\eta}{s} \sim \frac{l_{\mathrm{mfp}}p_{\mathrm{av}}}{k_{\mathrm{B}}} \propto l_{\mathrm{mfp}},
\label{eq.2}
\end{equation}
is found to be smaller for shorter mean free path $l_{\mathrm{mfp}}$. That is, a strongly interacting quantum fluid is a promising candidate to approach the KSS bound. 
\par
Because $\eta/s$ is related to the strength of particle-particle correlations, and Eq. (\ref{eq.1}) is independent of detailed system properties, $\eta/s$ has been considered as a useful quantity for the study of quantum fluids from the general viewpoint. For example, Sch{\"a}fer and Teaney evaluated $\eta/s$ from experimental data in some strongly correlated quantum fluids as\cite{Schafer:2009ab}
\begin{eqnarray}
\eta/s\ge 
\left\{
\begin{array}{ll}
8.8: & \mathrm{liquid}~^4\mathrm{He}, \\
6.3: & \mathrm{unitary~Fermi~atomic~gas~(^6Li)}, \\
5.0: & \mathrm{quark\mathchar`-gluon\mathchar`-plasma~(QGP)},
\end{array}
\right.
\label{eq.b}
\end{eqnarray}
in the unit of $\hbar/(4\pi k_{\mathrm{B}})$. Because $\eta/s\sim 380\times \hbar/(4\pi k_{\mathrm{B}})$ in water under the normal condition\cite{Kovtun:2005aa}, the minimum values of these quantum fluids in Eq. (\ref{eq.b}) are very close to the KSS bound.
\par
Although some theoretical counterexamples of the KSS conjecture have actually been pointed out\cite{Cohen:2007aa,Son:2008aa,Brigante:2008aa,Kats:2009aa,Buchel:2009aa,Rebhan:2012aa}, at least all the three real systems in Eq. (\ref{eq.b}) satisfy Eq. (\ref{eq.1}). However, it is still unclear whether or not one can obtain a smaller value of $\eta/s$ than Eq. (\ref{eq.b}) in some of these systems, when a more appropriate situation is considered. In this respect, an ultracold Fermi gas has an advantage, because a tunable interaction associated with a Feshbach resonance\cite{Chin:2010aa} allows us to investigate correlation effects on $\eta/s$ in a systematic manner.
\par
When we simply apply Eq. (\ref{eq.2}) to the BCS (Bardeen-Cooper-Schrieffer)-BEC (Bose-Einstein condensation) crossover regime\cite{Eagles:1969aa,Leggett:1980aa,Nozieres:1985aa,Sa-de-Melo:1993aa,Haussmann:1993aa,Haussmann:1994aa,Pistolesi:1994aa,Ohashi:2002aa} of an ultracold Fermi gas, we find that $\eta/s$ diverges in both the BCS and BEC limits, because the system becomes an ideal gas there ($l_{\mathrm{mfp}}\to\infty$). On the other hand, in the BCS-BEC crossover region, noting that $l_{\mathrm{mfp}}\sim 1/(n\sigma_s)$\cite{Pitaevskii:1981aa} (where $\sigma_s\propto a_s^2$ is the cross section with $a_s$ being the $s$-wave scattering length), one finds that Eq. (\ref{eq.2}) vanishes at the unitarity $a_s^{-1}=0$. Of course, this is a rough estimation; however, one may still expect that the unitary limit is a candidate for the interaction strength at which the lower bound of $\eta/s$ is obtained. This is one reason why $\eta$ and $\eta/s$ have recently attracted much attention around the unitary limit of an ultracold Fermi gas\cite{Bruun:2005aa,Massignan:2005aa,Cao:2011ab,Cao:2011aa,Elliott:2014aa,Elliott:2014ab,Punk:2006aa,Bruun:2007aa,Rupak:2007aa,Taylor:2010aa,Enss:2011aa,Guo:2011aa,Guo:2011ab,LeClair:2011aa,Salasnich:2011aa,Bruun:2012aa,Enss:2012ab,Goldberger:2012aa,Wlazlowski:2012aa,Chafin:2013aa,Romatschke:2013aa,Wlazlowski:2013aa,Bluhm:2014aa,Kryjevski:2014aa,Kikuchi:2016ab,Kagamihara:2017aa,Samanta:2017aa,Cai:2018aa,Joseph:2015aa,Wlazlowski:2015aa,Kagamihara:2019ab,Bluhm:2017aa}. 
\par
Although the condition for the lower bound of $\eta/s$ is still unknown in cold Fermi gas physics, the following recent studies may be helpful: (1) The observed $\eta$ in a $^6$Li Fermi gas takes a minimum value, not at the unitarity, but in the BEC side\cite{Elliott:2014ab}. This is consistent with the recent theoretical predictions\cite{Wlazlowski:2015aa,Kagamihara:2019ab}. Thus, the lower bound of $\eta/s$ might also be obtained in the BEC side. (2) Within the kinetic approach to a two-dimensional Fermi gas, Ref.\cite{Bruun:2012aa} recently predicted that mass imbalance lowers the magnitude of $\eta/s$. 
\par
Keeping (1) and (2) in mind, we theoretically assess the KSS conjecture in the normal state of a {\it three-dimensional} ultracold Fermi gas with mass imbalance. To evaluate the ratio $\eta/s$ in the BCS-BEC crossover region, we consistently include strong-coupling corrections to the shear viscosity $\eta$ and the entropy density $s$, within the framework of the self-consistent $T$-matrix approximation (SCTMA)\cite{Haussmann:1993aa,Haussmann:1994aa,Haussmann:2007aa,Enss:2011aa,Hanai:2014ab}. We then clarify how the ratio $\eta/s$ behaves in the phase diagram of a mass-imbalanced Fermi gas with respect to the temperature and the interaction strength. We also identify where the minimum of $\eta/s$ is obtained in this phase diagram, as well as the lower bound $(\eta/s)_{\mathrm{l.b.}}$, in the normal state above the superfluid phase transition temperature $T_{\mathrm{c}}$. We briefly note that SCTMA has been shown to (semi)quantitatively explain the observed $\eta$ and $s$ in a $^6$Li-$^6$Li unitary Fermi gas\cite{Elliott:2014ab,Joseph:2015aa,Haussmann:2007aa,Enss:2011aa,Kagamihara:2019ab}.
\par
We note that the KSS bound has also been extensively discussed in other research fields, such as high-energy QGP physics\cite{Song:2011aa}, condensed matter physics (graphene\cite{Muller:2009aa} and high-$T_{\mathrm{c}}$ cuprates\cite{Rameau:2014aa}), as well as liquid $^3$He and $^4$He\cite{Pakhira:2015aa}. Thus, clarifying the detailed condition to reach the lower bound of $\eta/s$ would make an impact on these fields. Regarding this, a tunable pairing interaction associated with a Feshbach resonance\cite{Chin:2010aa}, as well as the existence of various mass-imbalanced Fermi-Fermi mixtures, such as $^6$Li-$^{40}$K\cite{Wille:2008aa,Taglieber:2008aa,Trenkwalder:2011aa,Voigt:2009aa} and $^{40}$K-$^{161}$Dy\cite{Ravensbergen:2018aa,Ravensbergen:2018ab,Ravensbergen:2019aa}, would be great advantages of cold atom physics. 
\par
This paper is organized as follows: In Sec. \ref{sec:2}, we explain how to consistently evaluate $\eta$ and $s$ in the normal state of a mass-imbalanced Fermi gas, within the framework of SCTMA\cite{Haussmann:2007aa,Enss:2011aa,Kagamihara:2019ab,Hanai:2014ab}. In Sec. \ref{sec:3}, we show our numerical results on, not only $\eta/s$, but also $\eta$ and $s$, as well as effects of mass imbalance, in the BCS-BEC crossover region. In this paper, we set $\hbar = k_{\mathrm{B}} = 1$, and the system volume is taken to be unity, for simplicity.
\par
\par
\section{Formulation}
\label{sec:2}
\par
\begin{figure}[t]
\centering
\includegraphics[width=8cm]{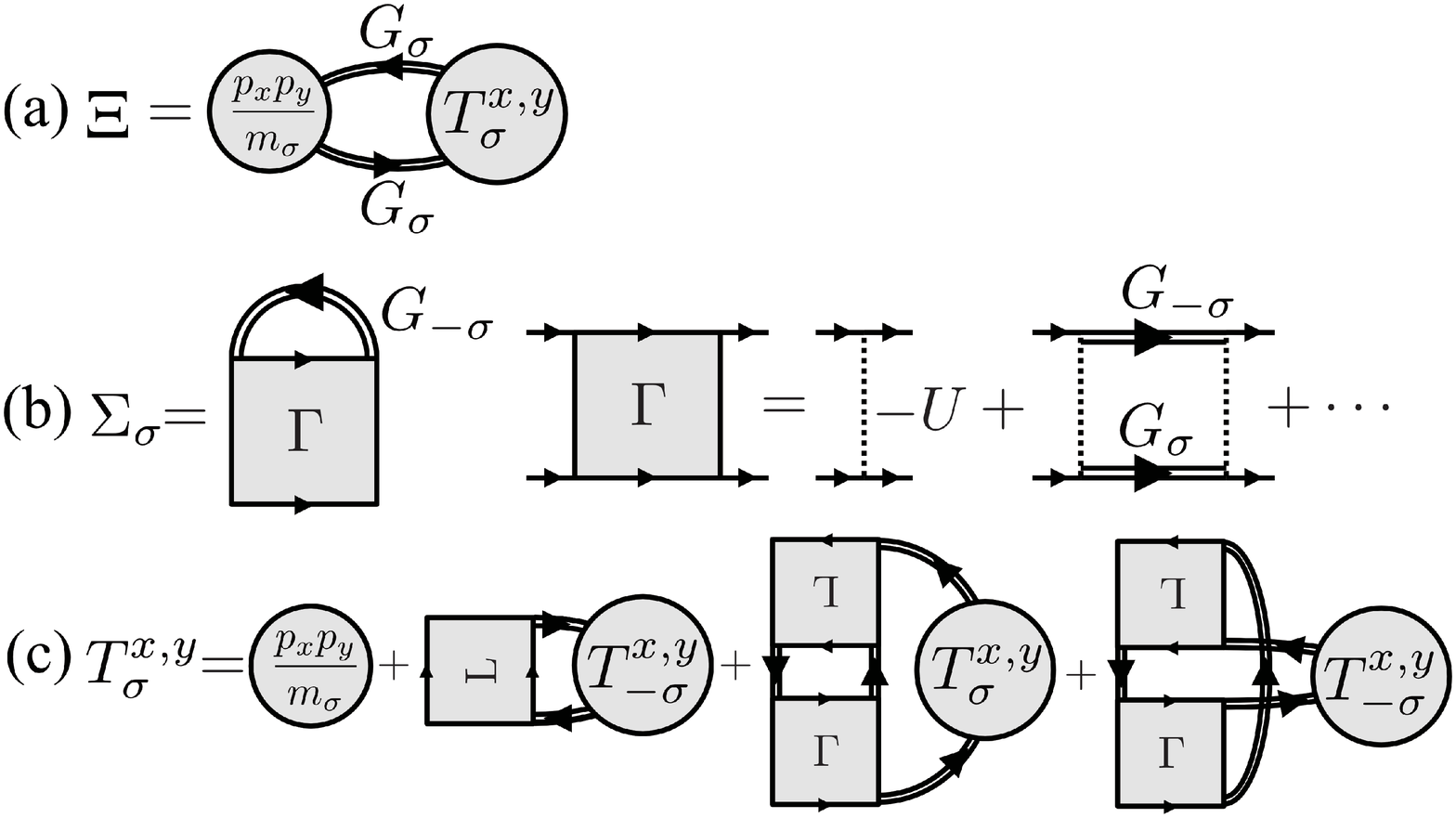}
\caption{(a) Diagrammatic representation for the shear-stress response function $\Xi(i\nu_m)$ in Eq. (\ref{eq.5}). The double solid line is the dressed Green's function $G_{\sigma}$, where $\sigma={\mathrm{L,H}}$ represent the light-mass (L) and heavy-mass (H) components. $p_xp_y/m_{\sigma}$ and $T^{x,y}_{\sigma}$ are the bare shear-stress vertex and the dressed one, respectively. (b) SCTMA self-energy $\Sigma_{\sigma}(\bm{p},i\omega_n)$ in the $\sigma$ component. The dashed line is a pairing interaction $-U$. (c) Dressed shear-stress vertex $T^{x,y}_{\sigma}$, being consistent with $\Sigma_{\sigma}(\bm{p},i\omega_n)$ in (b).
}
\label{fig1}
\end{figure}
\par
We consider a two-component Fermi gas with mass imbalance, described by the BCS-type Hamiltonian,
\begin{align} 
\nonumber
H&=
\sum_{\bm{p},\sigma=\mathrm{L,H}} \xi_{\bm{p},\sigma} c^{\dag}_{\bm{p},\sigma} c_{\bm{p},\sigma}
\\&
-U\sum_{\bm{p},\bm{p}',\bm{q}} c^{\dag}_{\bm{p}+\bm{q},\mathrm{L}} c^{\dag}_{\bm{p}'-\bm{q},\mathrm{H}} c_{\bm{p}',\mathrm{H}} c_{\bm{p},\mathrm{L}},
\label{eq.3}
\end{align}
where $c^\dagger_{\bm{p},\mathrm{L}}$ ($c^\dagger_{\bm{p},\mathrm{H}}$) is the creation operator of a Fermi atom with light mass $m_{\mathrm{L}}$ (heavy mass $m_{\mathrm{H}}$). $\xi_{\bm p,\sigma}=\varepsilon_{\bm p,\sigma}-\mu_{\sigma}=p^2/(2m_{\sigma})-\mu_{\sigma}$ is the kinetic energy of the $\sigma={\mathrm{L,H}}$ component, measured from the Fermi chemical potential $\mu_{\sigma}$. $-U~(<0)$ is a tunable pairing interaction associated with a Feshbach resonance\cite{Chin:2010aa}. 
\par
In the following two subsections, we explain how to calculate $\eta$ and $s$ in SCTMA.
\par
\par
\subsection{Shear viscosity}
\par
In the linear response theory\cite{Fetter:2003aa,Rickayzen:2013aa}, the shear viscosity $\eta$ is given by\cite{Kadanoff:1963aa,Luttinger:1964aa,Bruun:2007aa,Taylor:2010aa}
\begin{equation}
\eta = - \lim_{\omega \to 0}\frac{{\mathrm{Im}}[\Xi(\omega)]}{\omega}.
\label{eq.4}
\end{equation}
The shear-stress response function $\Xi(\omega)$ is obtained from the corresponding thermal correlation function,
\begin{align} \nonumber
\Xi(i\nu_m) &= T \sum_{\bp,\omega_n,\sigma} \frac{p_x p_y}{m_{\sigma}} G_{\sigma}(\bm{p},i\omega_n)T^{x,y}_{\sigma}(\bm{p},i\omega_n,i\omega_n+i\nu_m) 
\\
&\times 
G_{\sigma}(\bm{p},i\omega_n+i\nu_m),
\label{eq.5}
\end{align}
by way of the analytic continuation $i\nu_m\to\omega+i\delta$. Here, $\omega_n$ and $\nu_m$ are the fermion and boson Matsubara frequencies, respectively. $\delta$ is an infinitesimally small positive number. Equation (\ref{eq.5}) is diagrammatically described as Fig. \ref{fig1}(a), where $p_xp_y/m_{\sigma}$ is the bare shear-stress three-point vertex function, and $T^{x,y}_{\sigma}$ is the dressed one. We consistently treat the dressed vertex $T^{x,y}_{\sigma}$ and the self-energy $\Sigma_{\sigma}$ in the single-particle thermal Green's function,
\begin{equation}
G_{\sigma}(\bm{p},i\omega_n) = 
\frac{1}{i\omega_n - \xi_{\bm{p},\sigma} - \Sigma_{\sigma}(\bm{p},i\omega_n)},
\label{eq.6}
\end{equation}
so as to satisfy the Ward-Takahashi identity\cite{Schrieffer:1999aa,Baym:1961ab,Baym:1962aa,He:2014aa}. This is a required condition for any consistent theory. In SCTMA\cite{Haussmann:1993aa,Haussmann:2007aa,Enss:2011aa,Kagamihara:2019ab}, this condition is satisfied, by treating the diagrams in Figs. \ref{fig1}(b) and (c)\cite{Enss:2011aa,Kagamihara:2019ab} in a consistent manner. In these figures,
\begin{align}
\nonumber
\Gamma(\bm{q},i\nu_m)
&= 
-\frac{U}{1-U\Pi(\bm{q},i\nu_m)}
\\
&=
\frac{1}{\frac{m_{\mathrm{r}}}{4\pi a_s}+
\left[
\Pi(\bq,i\nu_m)
-\sum_{\bp} \frac{m_{\mathrm{r}}}{p^2}
\right]},
\label{eq.6b}
\end{align}
is the particle-particle scattering matrix, describing pairing fluctuations. Here,
\begin{equation}
\Pi(\bm{q},i\nu_m) = T \sum_{\bm{p},\omega_n} 
G_\mathrm{L}(\bm{p},i\omega_n) 
G_\mathrm{H}(\bm{q}-\bm{p},i\nu_m-i\omega_n),
\label{eq.7}
\end{equation}
is the pair correlation function. In Eq. (\ref{eq.6b}), we have absorbed the ultraviolet divergence involved in $\Pi({\bm q},i\nu_m)$ into the $s$-wave scattering $a_s$\cite{Haussmann:1993aa}, given by
\begin{equation}
\frac{4\pi \as}{m_{\mathrm{r}}}=
-\frac{U}{1-U\sum_{\bp}^{p_{\mathrm{c}}} \frac{m_{\mathrm{r}}}{p^2}},
\label{eq.8}
\end{equation}
where  $p_{\mathrm{c}}$ is a momentum cutoff and
\begin{equation}
m_{\mathrm{r}}= \frac{2m_{\mathrm{L}} m_{\mathrm{H}}}{m_{\mathrm{L}}+m_{\mathrm{H}}},
\label{eq.mr}
\end{equation}
equals twice the reduced mass. As usual, we measure the interaction strength in terms of the inverse scattering length $(k_{\mathrm{F}}a_s)^{-1}$, normalized by the Fermi momentum $k_{\mathrm{F}}$. In this scale, the weak-coupling BCS regime and strong-coupling BEC regime are characterized as $(k_{\mathrm{F}}a_s)^{-1} \lesssim -1$ and $(k_{\mathrm{F}}a_s)^{-1}\gtrsim +1$, respectively. The unitary limit is at $(k_{\mathrm{F}}a_s)^{-1}=0$.
\par
Using $\Gamma$ in Eq. (\ref{eq.6b}), we obtain $\Sigma_{\sigma}$ and $T^{x,y}_{\sigma}$ in SCTMA as, respectively, 
\begin{align}
&\Sigma_{\sigma}(p) = T \sum_{{\bm q}',\nu_m'} \Gamma(q') G_{-\sigma}(q'-p),
\label{eq.9}
\\ \nonumber
&T^{x,y}_{\sigma}(p,p+q)
\\ \nonumber
&=
\frac{p_xp_y}{m_{\sigma}}
+T\sum_{\bp',\omega_n'}
\Gamma(p+p'+q) \tilde{T}^{x,y}_{-\sigma}(p',p'+q)
\\ \nonumber
&-T\sum_{\bq',\nu_n'} G_{-\sigma}(q'-p) \Gamma(q') \Gamma(q'+q)
\\
&\times \sum_{\bp',\omega_n',\sigma'} \left[ G_{-\sigma'}(q'-p') \tilde{T}^{x,y}_{\sigma'}(p',p'+q) 
\right],
\label{eq.10}
\end{align}
where 
\begin{equation}
\tilde{T}^{x,y}_{\sigma}(p',p'+q) = G_{\sigma}(p')T^{x,y}_{\sigma}(p',p'+q) G_{\sigma}(p'+q), 
\label{eq.10b}
\end{equation}
and $-\sigma$ denotes the opposite component to $\sigma = \mathrm{L,H}$. In Eqs. (\ref{eq.10}) and (\ref{eq.10b}), we have used the abbreviated notations: $p=({\bm p},i\omega_n)$, $q=(0,\nu_m)$, $p'=({\bm p}',\omega_n')$, and $q'=({\bm q}',\nu_m')$. 
\par
In the weak-coupling BCS limit, the self-energy in Fig. \ref{fig1}(b), as well as the vertex correction in Fig. \ref{fig1}(c), can be ignored. The resulting $\Xi(i\nu_m)$ in Eq. (\ref{eq.5}) has the form,
\begin{align} \nonumber
\Xi(i\nu_m) &
\simeq T \sum_{\bp,\omega_n,\sigma} 
\frac{p_xp_y}{m_{\sigma}} G_{0,\sigma}(\bm{p},i\omega_n) \frac{p_xp_y}{m_{\sigma}}\
\\
&\times
G_{0,\sigma}(\bm{p},i\omega_n+i\nu_m),
\label{eq.11}
\end{align}
where 
\begin{equation}
G_{0,\sigma}({\bm p},i\omega_n)=
\frac{1}{i\omega_n-\xi_{\bm p,\sigma}},
\end{equation}
is the bare single-particle Green's function in the $\sigma$ component. Substituting the analytic continued Eq. (\ref{eq.11}) into Eq. (\ref{eq.4}), one obtains the diverging shear viscosity as,
\begin{equation}
\eta_{\mathrm{BCS~limit}}=
-\pi\sum_{{\bm p},\sigma}
\left(
\frac{p_xp_y}{m_\sigma}
\right)^2
\frac{\partial f(\xi_{\bm p,\sigma})}{\partial \xi_{\bm p,\sigma}}
\delta(0)\to\infty,
\label{eta_BCS}
\end{equation}
where $f(\xi_{\bm p,\sigma})$ is the Fermi distribution function.
\par
We also reach the same conclusion in the strong-coupling BEC limit: In this limit, the last two diagrams in Fig. \ref{fig1}(c) become dominant\cite{Kagamihara:2019ab}. In addition, $\Gamma$ in Eq. (\ref{eq.6b}) becomes proportional to the bare molecular Bose Green's function,
\begin{equation}
B_0({\bm q},i\nu_m)=
\frac{1}{i\nu_m-\xi_{\bm q}^{\mathrm{B}}},
\end{equation}
as\cite{Haussmann:1993aa}
\begin{equation}
\Gamma({\bm q},i\nu_m)=
\frac{8\pi}{m_{\mathrm{r}}^2a_s}
B_{0}({\bm q},i\nu_m).
\label{eq.12}
\end{equation}
Here, $\xi_{\bm q}^{\mathrm{B}}={\bm q}^2/(2M)-\mu_{\mathrm{B}}$ is the molecular kinetic energy, where $M=m_{\mathrm{L}} + m_{\mathrm{H}}$ is the molecular mass and $\mu_{\mathrm{B}}=2\mu+E_{\mathrm{bind}}$ is the Bose chemical potential, with $E_{\mathrm{bind}}=1/(m_{\mathrm{r}}a_s^2)$ being the binding energy of a two-body bound state. Then, $\Xi(i\nu_m)$ in the strong-coupling BEC limit has the form,
\begin{align} 
\nonumber
\Xi(i\nu_m) &
\simeq -T \sum_{\bq,\nu_m'} \frac{q_xq_y}{M} B_0(\bm{q},i\nu_m') \frac{q_xq_y}{M}
\\
&\times
B_0(\bm{q},i\nu_m'+i\nu_m).
\label{eq.13}
\end{align}
Using this, we obtain the shear viscosity in the strong-coupling BEC limit as,
\begin{equation}
\eta_{\mathrm{BEC~limit}}=
-\pi\sum_{\bm q}
\left(
\frac{q_xq_y}{M}
\right)^2
\frac{\partial n_{\mathrm{B}}(\xi_{\bm q}^{\mathrm{B}})}{\partial \xi_{\bm q}^{\mathrm{B}}}
\delta(0)\to\infty,
\label{eta_BEC}
\end{equation}
where $n_{\mathrm{B}}(\xi_{\bm q}^{\mathrm{B}})$ is the Bose distribution function. We briefly note that the diverging results in Eqs. (\ref{eta_BCS}) and (\ref{eta_BEC}) come from the infinite lifetime $\tau\to \infty$ of free Fermi atoms and free Bose molecules, respectively. (Note that $\eta\sim nl_{\mathrm{mfp}}p_{\mathrm{av}}\to\infty$, when $l_{\mathrm{mfp}}\propto\tau\to\infty$.)
\par
\par
\subsection{Entropy density}
\par
SCTMA satisfies the Tan's pressure relation\cite{Tan:2008aa,Tan:2008ac},
\begin{equation}
P = \frac{2}{3} E + \frac{C}{12\pi m_{\mathrm{r}} a_s},
\label{eq.14}
\end{equation}
where $P$, $E$, and $C$ are the pressure, the internal energy, and the Tan's contact, respectively. Substituting Eq. (\ref{eq.14}) into the thermodynamic identity for the entropy density $s$,
\begin{equation}
s= \frac{1}{T} \left[ P + E - \mu_{\mathrm{L}} N_{\mathrm{L}} - \mu_{\mathrm{H}} N_{\mathrm{H}} \right],
\end{equation}
we have
\begin{equation}
s = \frac{1}{T} \left[ \frac{5}{3} E + \frac{C}{12\pi m_{\mathrm{r}} a_s} - \mu_{\mathrm{L}} N_{\mathrm{L}} - \mu_{\mathrm{H}} N_{\mathrm{H}} \right].
\label{eq.15}
\end{equation}
Here, $N_\sigma$ is the number of atoms in the $\sigma={\mathrm{L,H}}$ component. The internal energy $E$ and the Tan's contact $C$ in SCTMA are given by, respectively,
\begin{equation}
E = T\sum_{\bm{p},\omega_n,\sigma} \left[ \frac{p^2}{2m_{\sigma}} + \frac{1}{2} \Sigma_{\sigma}(\bm{p},i\omega_n) \right] G_{\sigma}(\bm{p},i\omega_n),
\label{eq.16}
\end{equation}
\begin{equation}
C = -m_{\mathrm{r}}^2 T \sum_{\bm{q},\nu_m} \Gamma(\bm{q},i\nu_m).
\label{eq.17}
\end{equation}
\par
\begin{figure*}[t]
\centering
\includegraphics[width=14cm]{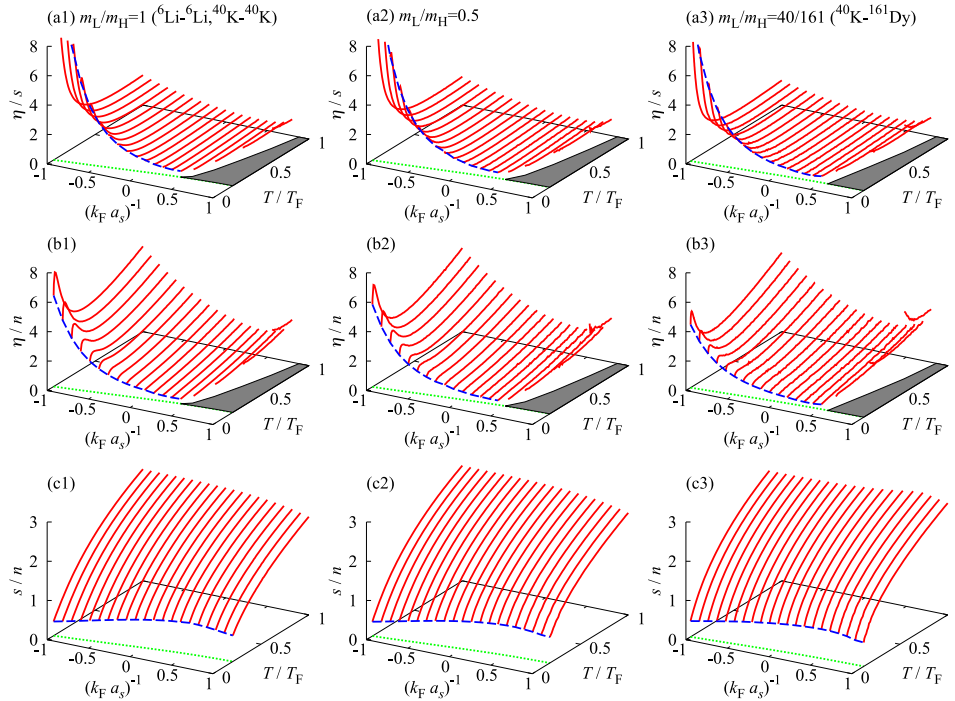}
\caption{(Color online) Calculated (a) $\eta/s$, (b) $\eta$, and (c) $s$, in the BCS-BEC crossover regime of an ultracold Fermi gas. Panels (i1), (i2), and (i3) (i=a,b,c) show the cases when $m_{\mathrm{L}}/m_{\mathrm{H}} = 1$ ($^6$Li-$^6$Li and $^{40}$K-$^{40}$K mixtures), $m_{\mathrm{L}}/m_{\mathrm{H}} = 0.5$, and $m_{\mathrm{L}}/m_{\mathrm{H}}=40/161$ ($^{40}$K-$^{161}$Dy mixture), respectively. The temperature is normalized by $T_{\mathrm{F}} = \kf^2 / (2m_{\mathrm{r}})$. $n$ is the total number density of Fermi atoms. The dotted line in the $T$-$(k_{\mathrm{F}}a_s)^{-1}$ plane is $T_{\mathrm{c}}$. The dashed line in each panel shows the result at $T_{\mathrm{c}}$. In panels (a) and (b), the gray-shaded areas in the BEC side near $T_{\mathrm{c}}$ are the regions where we could not calculate $\eta$ due to the numerical problem mentioned in the text.}
\label{fig2}
\end{figure*}
\begin{figure}[t]
\centering
\includegraphics[width=6cm]{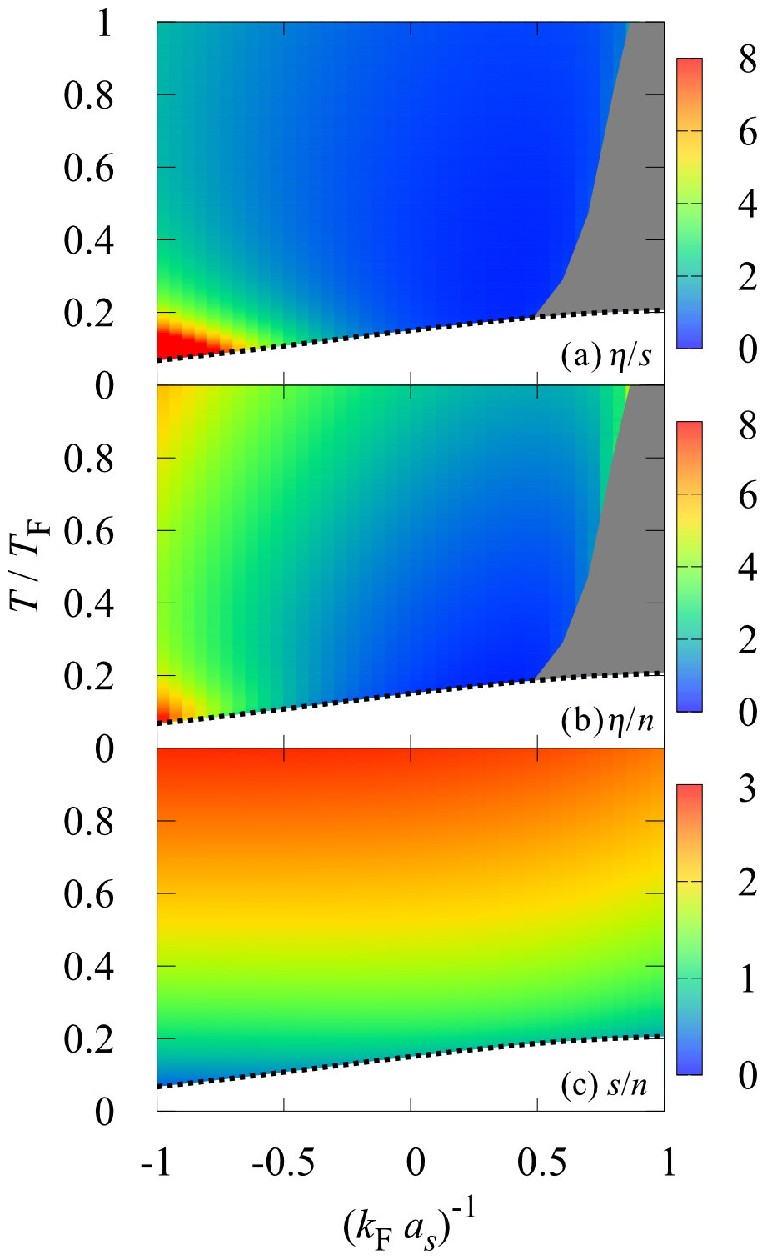}
\caption{(Color online) Same plots as Figs. \ref{fig2}(a1)-(c1), but using the density plots. In each panel, the dotted line shows $T_{\mathrm{c}}$. In panels (a) and (b), the gray-shaded areas in the BEC side are the regions where we could not calculate $\eta$ due to the numerical problem mentioned in the text.
}
\label{fig3}
\end{figure}
\par
\subsection{Calculations of $\mu_\sigma$ and $T_{\mathrm{c}}$, and computational note}
\par
In calculating $\eta$ and $s$, we need to evaluate $\mu_{\sigma}$, which is determined from the equation for the number $N_{\sigma}$ of Fermi atoms in the $\sigma$ component,
\begin{equation}
N_{\sigma} = T\sum_{\bm{p},\omega_n} G_{\sigma}(\bm{p},i\omega_n).
\label{eq.18}
\end{equation}
In this paper, we only deal with the population-balanced case ($N_{\mathrm{L}}=N_{\mathrm{H}}$). In this case, while the two components have different Fermi energies $\varepsilon_{{\mathrm{F}},\sigma={\mathrm{L,H}}}=k_{\mathrm{F}}^2/(2m_\sigma)$ and different Fermi temperatures $T_{{\mathrm{F}},\sigma={\mathrm{L,H}}}~(=\varepsilon_{{\mathrm{F}},\sigma}$), they have the common Fermi momentum $k_{\mathrm{F}}$. 
\par
At $T_{\mathrm{c}}$, we solve the number equation (\ref{eq.18}), together with the Thouless criterion\cite{Thouless:1960aa},
\begin{equation}
\Gamma^{-1}(\bm{q}=0,i\nu_m=0)=0,
\end{equation}
to determine $T_{\mathrm{c}}$ and $\mu_{\sigma}(T_{\mathrm{c}})$ in a consistent manner.
\par
We note that, while $s$ is directly obtained from Eq. (\ref{eq.15}), the analytic continuation $\Xi(\omega)=\Xi(i\nu_m\to\omega+i\delta)$ is needed to obtain $\eta$ in Eq. (\ref{eq.4}), which we numerically execute by the Pad\'e approximation\cite{Vidberg:1977aa}. Regarding this computation, as mentioned in our previous paper\cite{Kagamihara:2019ab}, we have sometimes faced the difficulty that the Pad\'e approximation unphysically gives negative or extraordinary large/small $\eta$, especially in the BEC regime near $T_{\mathrm{c}}$. We have also found that this problem depends on the detailed value of the cutoff momentum $k_{\mathrm{c}}$ in computing $\Xi(i\nu_m)$ in Eq. (\ref{eq.5}) (which is different from $p_{\mathrm{c}}$ in Eq. (\ref{eq.8})), as well as the number of Matsubara frequencies in executing the Pad\'e approximation. At this stage, we cannot completely avoid these problems. Thus, leaving these as our future problems, we take the same prescription as that in Ref.\cite{Kagamihara:2019ab}: We first numerically calculate $\Xi(i\nu_m)$ with various values of the momentum cutoff, $10k_{\mathrm{F}}\le k_{\mathrm{c}}\le 60k_{\mathrm{F}}$, in the momentum summation in Eq. (\ref{eq.5}). We then employ the Pad\'e approximation\cite{Vidberg:1977aa} to carry out the analytic continuation $i\nu_m\to \omega+i\delta$, retaining $50\sim 100$ Matsubara frequencies. We remove unphysical negative results from data, and also remove the highest and lowest 10\% of data to avoid the influence of abnormal results. We finally calculate the averaged value ${\bar \eta}$, as well as the standard deviation ${\bar \sigma}$ for the remaining data, to only retain the results satisfying $|\bar{\sigma}/\bar{\eta}|<0.1$. We show the regions where $|\bar{\sigma} / \bar{\eta}| \ge 0.1$ as the gray-shaded areas in Figs. \ref{fig2} and \ref{fig3}. $\bar{\sigma}$ will be used as the error of $\eta$ in Eqs. (\ref{eq.LB}) and (\ref{eq.KSS}), as well as in Figs. \ref{fig6}, \ref{fig8}, and \ref{fig9}.
\par
\par
\section{Results}
\label{sec:3}
\par
\par
\subsection{$\eta/s$ in the BCS-BEC crossover region}
\par
The uppermost three panels in Fig. \ref{fig2} show the calculated $\eta/s$ for various values of the mass-imbalance ratio $m_{\mathrm{L}}/m_{\mathrm{H}}$. For completeness, we also show SCTMA results for the shear viscosity $\eta$, as well as the entropy density $s$, in the lower panels. 
For clarity, we also show the density plots of Figs. \ref{fig2}(a1)-(c1) in Fig. \ref{fig3}. (As discussed soon later, since the overall structures of Figs. \ref{fig2}(a2)-(c2) and (a3)-(c3) are very similar to the mass-balanced case shown in panels (a1)-(c1), we only show the mass-balanced case in Fig. \ref{fig3}.)
When $m_{\mathrm{L}}\ne m_{\mathrm{H}}$, the heavy-mass component and light-mass component have different Fermi temperatures $T_{{\mathrm{F}},\sigma}=k_{\mathrm{F}}^2/(2m_\sigma)$, so that we scale the temperature $T$ in terms of the `averaged' Fermi temperature,
\begin{equation}
T_{\mathrm{F}}\equiv \frac{1}{2}[T_{\mathrm{F,H}}+T_{\mathrm{F,L}}]
=\frac{k_{\mathrm{F}}^2}{2m_{\mathrm{r}}},
\end{equation}
where $m_{\mathrm{r}}$ is given in Eq. (\ref{eq.mr}). 
\par
Figures \ref{fig2}(a1)-(a3) indicate that the overall behavior of $\eta/s$ as a function of the scaled interaction strength $(k_{\mathrm{F}}a_s)^{-1}$ and the scaled temperature $T/T_{\mathrm{F}}$ is almost the same among the three cases. We also find from the lower panels in Fig. \ref{fig2} that this result comes from the fact that $\eta$ and $s$ are also not so sensitive to the mass-imbalance ratio $m_{\mathrm{L}}/m_{\mathrm{H}}$. 
\par
To understand the dependence of $\eta/s$ on the temperature and the interaction strength shown in Figs. \ref{fig2}(a1)-(a3), it is useful to grasp the characteristic behavior of each $s$ and $\eta$: (1) The entropy density $s$ always monotonically decreases with decreasing the temperature, and is not so sensitive to the interaction strength (see panels (c1)-(c3)). (2) In addition to the diverging behavior in the BCS and BEC limits (see Eqs. (\ref{eta_BCS}) and (\ref{eta_BEC})), the shear viscosity $\eta$ exhibits non-monotonic temperature dependence. In the mass-balanced case (panel (b1)), the origin of this non-monotonic behavior has been explained in Ref.\cite{Kagamihara:2019ab}. Because of the above-mentioned similarity among Figs. \ref{fig2}(b1)-(b3), we expect that this explanation would also be valid for the mass-imbalanced case. That is, (i) at high temperatures, the temperature dependence of $\eta\propto T^\kappa$ (where $\kappa$ is a positive constant) may be understood as a property of an ordinary classical gas\cite{Bruun:2005aa,Massignan:2005aa}. (ii) In the weak-coupling regime at low temperatures ($T\ll T_{\mathrm{F}}$), Fermi quasi-particle scatterings are suppressed by the Pauli blocking, which elongates the mean free path as $l_{\mathrm{mfp}}\propto T^{-2}$, leading to $\eta\propto T^{-2}$ (see Eq. (\ref{eq_simple})). (iii) In the BCS side near $T_{\mathrm{c}}$, the enhancement of pairing fluctuations again shortens $l_{\mathrm{mfp}}$, giving the decrease of $\eta$ with decreasing the temperature\cite{Kagamihara:2019ab}. This makes a peak structure near $T_{\mathrm{c}}$ in this regime (see Figs. \ref{fig2}(b1)-(b3)). (iv) In the strong-coupling BEC regime, when Fermi atoms form stable molecules overwhelming thermal dissociation, $\eta$ increases with decreasing the temperature, reflecting the increase of the weakly interacting stable bosons.
\par
\begin{figure}[t]
\centering
\includegraphics[width=6cm]{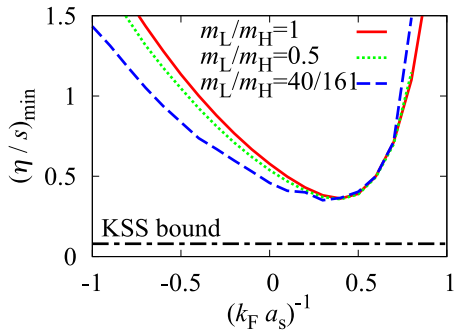}
\caption{(Color online) Evaluated minimum value $(\eta/s)_{\mathrm{min}}$ as a function of the interaction strength $(k_{\mathrm{F}}a_s)^{-1}$.}
\label{fig4}
\end{figure}
\par
From the above discussions on $s$ and $\eta$, one finds that the interaction dependence of $\eta/s$ is mainly determined by $\eta$. In particular, $\eta/s$ diverges in the BCS and BEC limits, because of the diverging $\eta$ there. For the temperature dependence, since $s$ always monotonically decreases with decreasing temperatures, $\eta/s$ increases with decreasing the temperature in the BCS and the BEC regime near $T_{\mathrm{c}}$ where $\eta$ increases with decreasing the temperature\cite{note}. In the high-temperature region where both $s$ and $\eta$ decrease as one decreases the temperature, one cannot immediately conclude the detailed temperature dependence of $\eta/s$. However, our numerical results show that it always decreases with decreasing the temperature in this regime. As a result, for a given interaction strength, the temperature dependence of $\eta/s$ always takes a minimum value ($\equiv(\eta/s)_{\mathrm{min}}$) at a certain temperature ($\equiv T_{\mathrm{min}}$) above $T_{\mathrm{c}}$, as shown in Fig. \ref{fig4}. 
\par
\begin{figure*}[t]
\centering
\includegraphics[width=13.6cm]{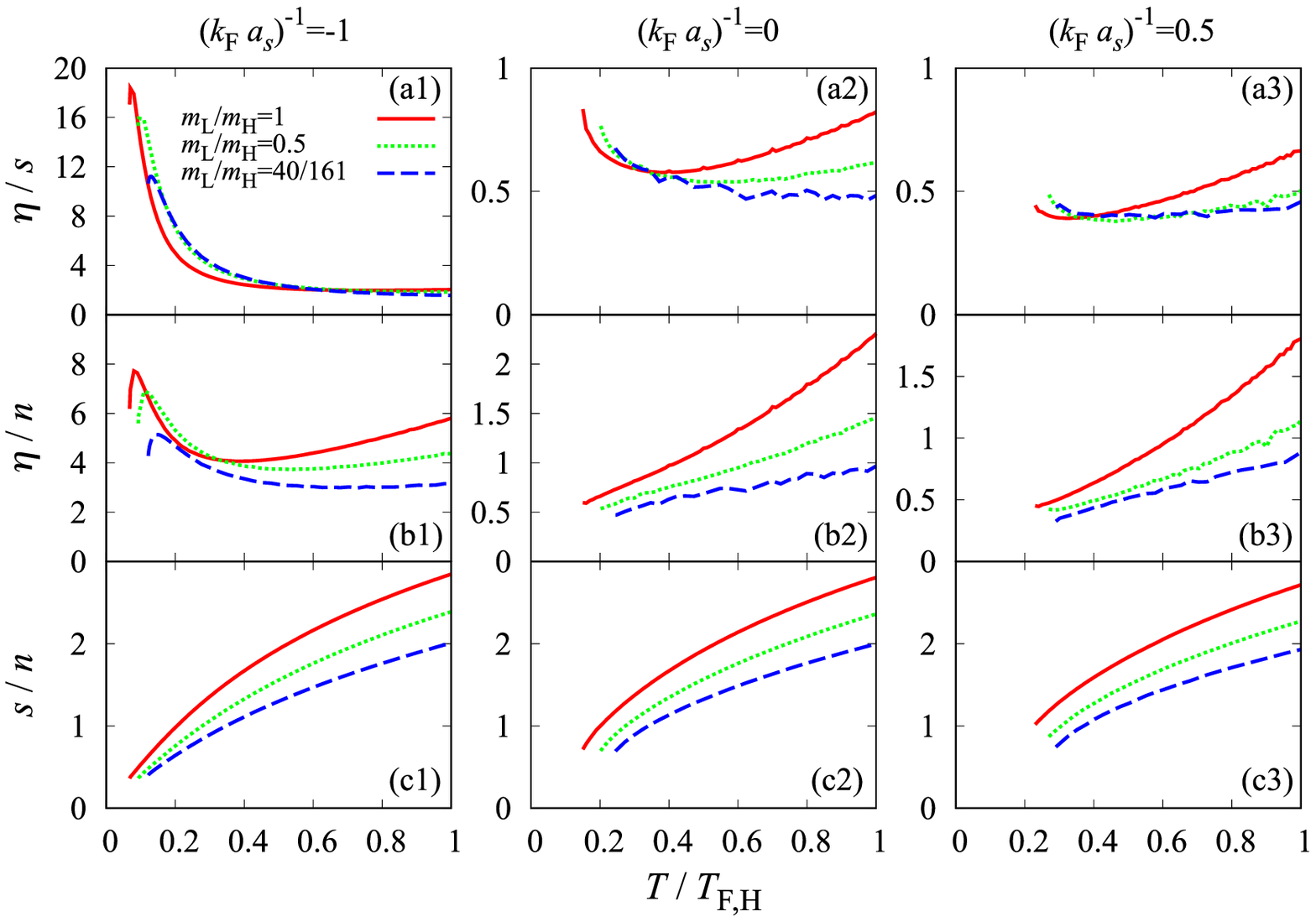}
\caption{(Color online) Calculated temperature dependence of (a) $\eta/s$, (b) $\eta$, and (c) $s$, in the normal state above $T_{\mathrm{c}}$. Panels (i1), (i2), and (i3) (i=a,b,c), show the cases when $(\kf\as)^{-1}=-1$ (BCS side), $(\kf\as)^{-1}=0$ (unitarity limit), and  $(\kf\as)^{-1}=0.5$ (BEC side), respectively. The temperature is normalized by the Fermi temperature $T_{\mathrm{F},{\mathrm{H}}} = \kf^2/(2m_{\mathrm{H}})$ of the heavy-mass component. The values of $T_{\mathrm{min}}$ in panels (a1)-(a3) are as follows: (a1) $T_{\mathrm{min}}/T_{\mathrm{F,H}}=0.78$ ($m_{\mathrm{L}}/m_{\mathrm{H}}=1$), 1.03 ($m_{\mathrm{L}}/m_{\mathrm{H}}=0.5$), and 1.43 ($m_{\mathrm{L}}/m_{\mathrm{H}}=40/161$). (a2) $T_{\mathrm{min}}/T_{\mathrm{F,H}}=0.38$ ($m_{\mathrm{L}}/m_{\mathrm{H}}=1$), 0.538 ($m_{\mathrm{L}}/m_{\mathrm{H}}=0.5$), and 0.974 ($m_{\mathrm{L}}/m_{\mathrm{H}}=40/161$). (a3) $T_{\mathrm{min}}/T_{\mathrm{F,H}}=0.33$ ($m_{\mathrm{L}}/m_{\mathrm{H}}=1$), 0.46 ($m_{\mathrm{L}}/m_{\mathrm{H}}=0.5$), and 0.574 ($m_{\mathrm{L}}/m_{\mathrm{H}}=40/161$).
}
\label{fig5}
\end{figure*}
\par
\subsection{Lower bound of $\eta/s$ and effects of mass imbalance}
\par
Although Figs. \ref{fig2}(a1)-(a3) look similar to one another, Fig. \ref{fig4} shows that the evaluated $(\eta/s)_{\mathrm{min}}$ actually depends on $m_{\mathrm{L}}/m_{\mathrm{H}}$. We here discuss detailed mass-imbalance effects on $\eta/s$, especially near the lower bound of this ratio.
\par
Figure \ref{fig5} shows $\eta/s$, $\eta$, and $s$, as functions of the temperature. Here, for later convenience, the temperature is normalized by the Fermi temperature $T_{\mathrm{F,H}}$ of the heavy-mass component. This corresponds to the situation that we fix $m_{\mathrm{H}}$ and tune the mass-imbalance ratio $m_{\mathrm{L}}/m_{\mathrm{H}}$ by adjusting (decreasing) $m_{\mathrm{L}}$. We briefly note that this scaled temperature $T/T_{\mathrm{F,H}}$ is also used in Ref.\cite{Bruun:2012aa} in examining $\eta$ in a mass-imbalanced Fermi gas by the kinetic theory.
\par
We see in Figs. \ref{fig5}(b1)-(b3) that the mass imbalance ($m_{\mathrm{L}}/m_{\mathrm{H}}<1$) suppresses $\eta$, except in the weak-coupling case shown in panel (b1). Although this result is consistent with the previous work\cite{Bruun:2012aa}, we find in panels (c1)-(c3) that the mass imbalance also suppresses $s$. (We will discuss background physics of these mass-imbalance effects on $\eta$ and $s$ in Sec. 3.3.) As a result, the effects of mass imbalance on $\eta/s$ around $T_{\mathrm{min}}$ are not so remarkable, compared to the suppression of $\eta$ (see Figs. \ref{fig5}(a1)-(a3)). 
\par
\begin{figure}[t]
\centering
\includegraphics[width=7cm]{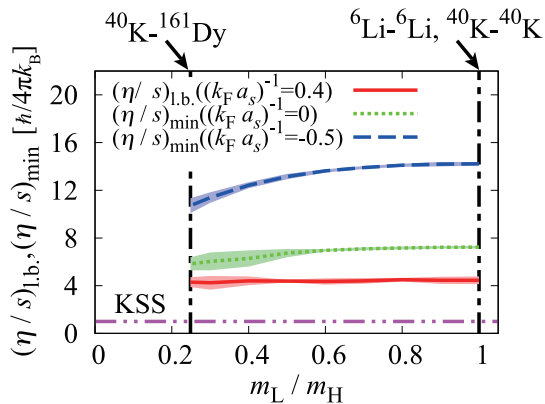}
\caption{(Color online) Calculated lower bound $(\eta/s)_{\mathrm{l.b.}}$ in the unit of $\hbar/(4\pi k_{\mathrm{B}})$, which is almost independent of $m_{\mathrm{L}}/m_{\mathrm{H}}~(\le 1)$. This lower bound is obtained at $(\kf\as)^{-1}\simeq 0.4$ near $T_{\mathrm{c}}$. For comparison, we also plot $(\eta/s)_{\mathrm{min}}$ at $(\kf\as)^{-1}=-0.5$ and at $(\kf\as)^{-1}=0$. The error bars show $3{\bar \sigma}/s$. The same error bars are also used in Figs. \ref{fig8}(a) and \ref{fig9}.
Each vertical line corresponds to the Fermi-Fermi mixture written above this figure. `KSS' shows the KSS bound in Eq. (\ref{eq.1}).
}
\label{fig6}
\end{figure}

In particular, as shown in Fig. \ref{fig6}, the lower bound of $\eta/s$ ($\equiv (\eta/s)_{\mathrm{l.b.}}$), which is given as the minimum value of $(\eta/s)_{\mathrm{min}}$, is almost $m_{\mathrm{L}}/m_{\mathrm{H}}$-independent, although, except for this special case, $(\eta/s)_{\mathrm{min}}$ depends on $m_{\mathrm{L}}/m_{\mathrm{H}}$. At present, we can only deal with the case down to $m_{\mathrm{L}}/m_{\mathrm{H}}=40/161 \simeq 0.248$ ($^{40}$K-$^{161}$Dy Fermi mixture) because of the computational problem. At least within our numerical results, the lowest value $(\eta/s)_{\mathrm{l.b.}}$ seems almost universal in a mass-imbalanced Fermi gas. 
\par
\par
Comparing this calculated lower bound with the KSS bound in Eq. (\ref{eq.1}), one finds
\begin{align}
\left( \frac{\eta}{s} \right)_{\mathrm{l.b.}} = (4.5 \pm 0.2) \times \frac{\hbar}{4\pi k_{\mathrm{B}}},
\label{eq.LB}
\end{align}
where the error is the maximum value of $\bar{\sigma}/s$ among the calculated cases, $40/161\le m_{\mathrm{L}}/m_{\mathrm{H}}\le 1$. $\bar{\sigma}$ is defined at the end of Sec. \ref{sec:2}. As seen in Fig. \ref{fig4}, $\eta/s$ reaches this lower bound, when $(k_{\mathrm{F}}a_s)^{-1}\simeq 0.4$. At this stage, we cannot analytically prove the universality of the lower bound in Eq. (\ref{eq.LB}) in a mass-imbalanced Fermi gas. However, at least numerically, our result indicates that Eq. (\ref{eq.LB}) holds in the wide parameter region with respect to the mass-imbalance ratio. Thus, it would be an interesting experimental challenge to confirm this prediction by observing $^6$Li-$^6$Li\cite{Bluhm:2017aa}, $^{40}$K-$^{40}$K\cite{Jin}, and $^{40}$K-$^{161}$Dy\cite{Ravensbergen:2018aa,Ravensbergen:2018ab,Ravensbergen:2019aa} mixtures. At present, our theory cannot examine the more highly imbalanced $^6$Li-$^{40}$K mixture\cite{Wille:2008aa,Taglieber:2008aa,Trenkwalder:2011aa,Voigt:2009aa} ($m_{\mathrm{L}}/m_{\mathrm{H}}=6/40=0.15$) because of the computational problem. However, it would be interesting to experimentally examine whether or not this universality still holds even there. We briefly note that the tuning of the interaction strength by using a Feshbach resonance is possible in all these Fermi gases.
\par
\par
\subsection{Mass-imbalance effects on $\eta$ and $s$ in the BCS-BEC crossover region}
\par
Although the main topic of this paper is $\eta/s$, we here discuss mass-imbalance effects on $\eta$ and $s$ in the BCS-BCS crossover region.
\par
\subsubsection{Shear viscosity $\eta$}
\par
In the weak-coupling BCS regime, when the temperature $T$ is slightly below $T_{\mathrm{F,L}}$ but is still much higher than $T_{\mathrm{F,H}}$, while the light-mass component enters the Fermi degenerate regime, the heavy-mass component is still in the classical regime. Thus, compared to the mass-balanced case near the Fermi temperature, the suppression of quasi-particle scatterings by the Pauli blocking would be weak. This shortens the mean free path $l_{\mathrm{mfp}}$, which decreases $\eta$\cite{Bruun:2012aa}. This tendency can be seen in Fig. \ref{fig5}(b1) at high temperatures.
\par
When $T\lesssim T_{\mathrm{F,H}}$ in the BCS regime, both the light-mass and heavy-mass components are in the Fermi degenerate regime. In this case, the light-mass component is already deep inside the Fermi degenerate regime ($T\ll T_{\mathrm{F,L}}$). As a result, the Pauli blocking effect on quasi-particle scatterings works more remarkably, compared to the case of a mass-balanced Fermi gas slightly below the Fermi temperature. Thus, the upturn behavior of $\eta$ (which originates from the suppression of quasi-particle scatterings by the Fermi degeneracy) starts to occur from a higher temperature in the mass-imbalanced case than in the mass-balanced case. Indeed, Fig. \ref{fig5}(b1) shows that the temperature at which $\eta$ takes a minimum value increases, as the mass-imbalance ratio $m_{\mathrm{L}}/m_{\mathrm{H}}$ decreases from unity. 
\par
In the strong-coupling BEC regime, the Pauli blocking effect is no longer expected. Instead, because system properties are dominated by tightly bound molecules in this regime, their correlations play crucial roles for the mass-imbalance effects on $\eta$: In SCTMA, an effective molecular interaction $U_{\mathrm{B}}$ in this regime has the form\cite{Haussmann:1993aa},
\begin{align}
U_{\mathrm{B}} = \frac{4\pi a_{\mathrm{B}}}{M},
\end{align}
where $M=m_{\mathrm{L}}+m_{\mathrm{H}}$ is the molecular mass, and 
\begin{align}
a_{\mathrm{B}} = \frac{M}{m_{\mathrm{r}}}\as = \frac{(m_{\mathrm{L}}+m_{\mathrm{H}})^2}{2m_{\mathrm{L}} m_{\mathrm{H}}} \as.
\end{align}
is the $s$-wave molecular scattering length, which increases as $m_{\mathrm{L}}/m_{\mathrm{H}}$ decreases from unity. This mass-imbalance effect thus decreases $\eta$.
\par
In addition, the so-called intraband scattering process is known to contribute to the molecular damping $\gamma(\bq,\omega\geq 0)$, which is given by\cite{Kagamihara:2019ab},
\begin{align}
\gamma(\bq,\omega\geq 0) 
\sim 
\sum_{\sigma} \frac{m_{\sigma}^2\Delta_{\mathrm{pg}}^2}{16\pi|\mu_{\mathrm{av}}|^2} e^{\mu_{\sigma}/T} \left(\frac{\omega}{q}\right)
e^{-\frac{m_{\sigma}}{2T} \left(\frac{\omega}{q}\right)^2}.
\label{eq:eq28}
\end{align}
Here, $\mu_{\mathrm{av}}=(\mu_{\mathrm{L}}+\mu_{\mathrm{H}})/2$ is the averaged chemical potential, and $\Delta_{\mathrm{pg}}=\sqrt{-T\sum_{{\bm q},\nu_m}\Gamma({\bm q},i\nu_m)}$ is sometimes referred to as the pseudogap parameter in the literature, which physically describes effects of pairing fluctuations. In the strong-coupling BEC regime, both $\mu_{\mathrm{L}}$ and $\mu_{\mathrm{H}}$ are negative, satisfying $|\mu_{\mathrm{L}}+\mu_{\mathrm{H}}| = E_{\mathrm{bind}}$ (where $E_{\mathrm{bind}}=1/(m_{\mathrm{r}}a_s^2)$ is the binding energy of a two-body bound state). In the presence of mass imbalance, one finds that $\mu_{\mathrm{L}} > \mu_{\mathrm{H}}$, and $\mu_{\mathrm{L}} - \mu_{\mathrm{H}}$ becomes large when the mass-imbalance ratio $m_{\mathrm{L}}/m_{\mathrm{H}}$ decreases from unity. This enhances the molecular damping $\gamma(\bq,\omega\geq 0)$ in Eq. (\ref{eq:eq28}), leading to the suppression of $\eta$, as seen in Fig. \ref{fig5}(b3).
\par
\subsubsection{Entropy density $s$}
\par
In the weak-coupling BCS regime, both the $\sigma={\mathrm{L,H}}$ components start to enter the Fermi degenerate regime when $T\lesssim T_{\mathrm{F,H}}$. In this case, the light-mass component is already deep inside this quantum regime ($T\ll T_{\mathrm{F,L}}$), so that $s$ becomes smaller than that in the mass-balanced case near the Fermi temperature (see Fig. \ref{fig5}(c1)). 
\par
In the BEC regime, (1) when $m_{\mathrm{L}}$ decreases with fixing $m_{\mathrm{H}}$, molecular mass $M = m_{\mathrm{L}} + m_{\mathrm{H}}$ becomes light, which raises the BEC transition temperature $T_{\mathrm{BEC}}$. (2) The entropy density $s=5\zeta(5/2)/[4\zeta(3/2)] n \simeq 0.642n$ of an ideal Bose gas with $n/2$ molecular density at $T_{\mathrm{BEC}}$ does not depend on mass (where $n$ is the total number density of Fermi atoms, and $\zeta(x)$ is the Riemann's zeta function). (3) Above $T_{\mathrm{BEC}}$, lighter molecular mass gives smaller entropy density, because thermal excitations of lighter-mass particles are more difficult. Because of (1)-(3), $s$ becomes small as $m_{\mathrm{L}}/m_{\mathrm{H}}$ decreases from unity, as seen in Fig. \ref{fig5}(c3).
\par
\par
\subsection{Key quantum phenomena for the lower bound of $\eta/s$}
\label{sec:4}
\par
\begin{figure}[t]
\centering
\includegraphics[width=6cm]{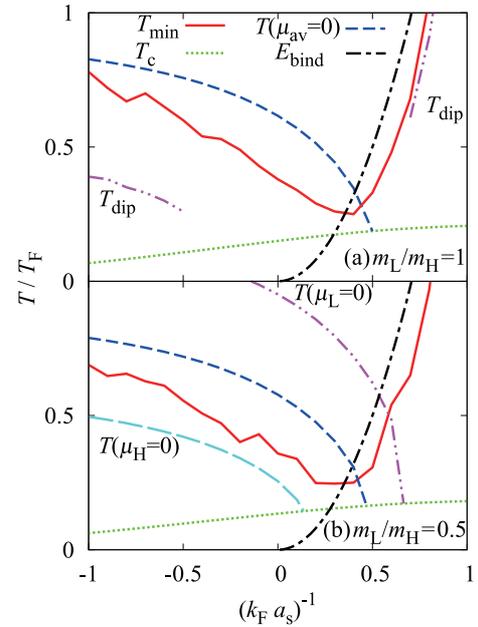}
\caption{(Color online) (a) Temperature $T_{\mathrm{min}}$ at which $(\eta/s)_{\mathrm{min}}$ is obtained in the mass-balanced case ($m_{\mathrm{L}}/m_{\mathrm{H}}=1$). $T(\mu_{\mathrm{av}}=0)$ is the temperature at which the Fermi chemical potential $\mu~(=\mu_{\mathrm{H}}=\mu_{\mathrm{L}})$ vanishes. $E_{\mathrm{bind}}=1/(m a_s^2)$ is the binding energy of a two-body bound state (where $m=m_{\mathrm{H}}=m_{\mathrm{L}}$ is the atomic mass). $T_{\mathrm{dip}}$ is the temperature at which $\eta(T)$ exhibits a dip. (b) $T_{\mathrm{min}}$ and $T(\mu_{\mathrm{av}}=0)$ in the mass-imbalanced case when $m_{\mathrm{L}}/m_{\mathrm{H}}=0.5$. We also plot $T(\mu_{\mathrm{H}}=0)$ and $T(\mu_{\mathrm{L}}=0)$ below which $\mu_{\mathrm{H}}$ and $\mu_{\mathrm{L}}$ become positive, respectively.
}
\label{fig7}
\end{figure}
\par
Figure \ref{fig7} shows the temperature $T_{\mathrm{min}}$ (where $(\eta/s)_{\mathrm{min}}$ is obtained at each interaction strength). We find from this figure that the lowest $T_{\mathrm{min}}$ is realized when $(\kf \as)^{-1} \simeq 0.4$ in both the mass-balanced and mass-imbalanced cases shown in panels (a) and (b), respectively. Here, we point out that the lower bound $(\eta/s)_{\mathrm{l.b.}}$ in Eq. (\ref{eq.LB}) is also obtained at this interaction strength (see Fig. \ref{fig4}). Using this fact, we explain background physics of $(\eta/s)_{\mathrm{l.b.}}$ in this subsection. 
\par
\subsubsection{Origin of the interaction dependence of $T_{\mathrm{min}}$}
\par
We first discuss the origin of the interaction dependence of $T_{\mathrm{min}}$ in the mass-balanced case shown in Fig. \ref{fig7}(a). For this purpose, we introduce the two characteristic temperatures (1) $T(\mu_{\mathrm{av}}=0)$ (below which $\mu_{\mathrm{av}}>0$) and (2) $T=E_{\mathrm{bind}}$. Then, the behavior of $T_{\mathrm{min}}$ in Fig. \ref{fig7}(a) can be explained as follows:
\par
\begin{enumerate}
\item[(i)] 
In the weak-coupling BCS regime, while $\eta$ decreases with decreasing the temperature in the classical regime\cite{Bruun:2005aa,Massignan:2005aa,Bruun:2012aa}, it exhibits a dip structure at a certain temperature ($\equiv T_{\mathrm{dip}}$) in the Fermi degenerate regime, below which $\eta$ increases with decreasing the temperature. This is because the Pauli blocking brings about a long quasi-particle lifetime, leading to the enhancement of $\eta$. Although $\eta$ again decreases near $T_{\mathrm{c}}$ due to pairing fluctuations (see Fig. \ref{fig2}(b1))\cite{Kagamihara:2019ab}, $\eta/s$ is still very large there because of small $s$. Thus, in the weak-coupling BCS regime, the temperature $T_{\mathrm{min}}$ at which $\eta/s$ becomes minimum is related to $T_{\mathrm{dip}}$. Since the Pauli blocking is a Fermi-surface effect, it only works in the Fermi degenerate regime below the characteristic temperature $T(\mu_{\mathrm{av}}=0)$ which is defined as the temperature below which $\mu_{\mathrm{av}}$ becomes positive. That is, $T_{\mathrm{dip}}<T(\mu_{\mathrm{av}}=0)$ in the BCS regime.
\par
In the BCS regime, the monotonic temperature dependence of $s$ in the ratio $\eta/s$ simply gives $T_{\mathrm{min}}>T_{\mathrm{dip}}$. On the other hand, because the interaction dependence of $s$ is weak (see Fig. \ref{fig2}(c1)  and \ref{fig3}(c)), the interaction dependence of $T_{\mathrm{min}}$ is similar to that of $T(\mu_{\mathrm{av}}=0)$ and $T_{\mathrm{dip}}$, as shown in Fig. \ref{fig7}(a). As one approaches the BCS-BEC crossover region from the weak-coupling side, $T(\mu_{\mathrm{av}}=0)$ eventually reaches $T_{\mathrm{c}}$ at $(k_{\mathrm{F}}a_s)^{-1}\simeq 0.5$, so that $T_{\mathrm{min}}$ also becomes close to $T_{\mathrm{c}}$ there. In the unitary regime, a strong pairing interaction gives a short lifetime of Fermi quasi-particles, giving small $\eta/s$ there. 
\par
\item[(ii)] 
In the strong-coupling BEC regime, when the temperature becomes lower than $T=E_{\mathrm{bind}}$, the formation of two-body bound molecules starts to occur, overwhelming their thermal dissociations. In such a molecular Bose gas, the large binding energy $E_{\mathrm{bind}}=1/(m_{\mathrm{r}}a_s^2)$ and weak molecular-scattering effects\cite{Haussmann:1993aa,Haussmann:1994aa,Kagamihara:2019ab} lead to a long lifetime $\tau_{\mathrm{B}}$ of Bose quasi-particles, giving large $\eta\propto \tau_{\mathrm{B}}$. Thus, as one decreases the temperature from $T\gg E_{\mathrm{bind}}$, $\eta$ exhibits a dip at $T_{\mathrm{dip}}\sim E_{\mathrm{bind}}$. As a result, $T_{\mathrm{min}}$ is located close to $E_{\mathrm{bind}}$ (see the BEC regime in Fig. \ref{fig7}(a)). Because of the weak interaction dependence of $s$ in the BEC side (see Figs. \ref{fig2}(c1) and \ref{fig3}(c)), the interaction dependence of $T_{\mathrm{min}}$ is similar to that of $E_{\mathrm{bind}}$. Then, since $E_{\mathrm{bind}}=1/(m_{\mathrm{r}} a_s^2)$ vanishes at the unitarity $(k_{\mathrm{F}}a_s)^{-1}=0$, $T_{\mathrm{min}}$ decreases with approaching the unitary limit from the BEC side.
\end{enumerate}
\par
Together with the above discussions starting from the BCS regime (i) and BEC regime (ii), one reaches the conclusion that $T_{\min}$ takes the lowest value at
\begin{equation}
T(\mu_{\mathrm{av}}=0)\simeq E_{\mathrm{bind}}.
\label{eq.aa}
\end{equation}
Indeed, Fig. \ref{fig7}(a) shows that Eq. (\ref{eq.aa}) is realized at $(k_{\mathrm{F}}a_s)^{-1}\simeq 0.4$.
\par
The above discussion can be extended to the mass-imbalanced case shown in Fig. \ref{fig7}(b): In this case, the temperature ($\equiv T(\mu_{\mathrm{L}}=0)$) below which $\mu_{\mathrm{L}}>0$ is different from the temperature ($\equiv T(\mu_{\mathrm{H}}=0)$) below which $\mu_{\mathrm{H}}>0$ (see Fig. \ref{fig7}(b)). However, even in this situation, we still expect that the Pauli blocking effect would become crucial below the above-mentioned characteristic temperature $T(\mu_{\mathrm{av}}=0)$ (below which the averaged chemical potential $\mu_{\mathrm{av}}=[\mu_{\mathrm{L}}+\mu_{\mathrm{H}}]/2$ becomes positive). Indeed, Fig. \ref{fig7}(b) shows that, in the mass-imbalanced case, the interaction dependence of $T_{\mathrm{min}}$ in the BCS side is still close to that of $T(\mu_{\mathrm{av}}=0)$ as in the mass-balanced case. Because $T=E_{\mathrm{bind}}$ does not depend on $m_{\mathrm{L}}/m_{\mathrm{H}}$ in Fig. \ref{fig7}, the above discussion for a mass-balanced Fermi gas is also valid for the mass-imbalanced case.  
\par
\subsubsection{Lower bound of $\eta/s$ and two quantum effects}
\par
Because (1) $\eta/s$ becomes small in the unitary regime due to short atomic mean free path $l_{\mathrm{mfp}}$ (see Eq. (\ref{eq.2})), and (2) $\eta/s$ decreases with decreasing the temperature when $T\ge T_{\min}$, $(\eta/s)_{\mathrm{l.b.}}$ is obtained at $(k_{\mathrm{F}}a_s)^{-1}\simeq 0.4$ where $T_{\mathrm{min}}$ takes the lowest value. 
\par
As mentioned in the introduction, the appearance of $\hbar$ in the right-hand side of Eq. (\ref{eq.1}) means that the KSS bound is associated with a quantum phenomenon. To understand this, $T_{\mathrm{min}}$ is also useful: In the weak-coupling BCS regime, the Pauli blocking effect occurring in the Fermi degenerate regime (which determines $T_{\mathrm{min}}$ in this regime) is just a {\it quantum statistical phenomenon}. In the BEC side, we also recall that the formation of the two-body bound state is a {\it quantum mechanical phenomenon}, because the binding energy involves $\hbar$ as $E_{\mathrm{bind}} = \hbar^2 / m_{\mathrm{r}} a_s^2$. Thus, at least in an ultracold Fermi gas, these two quantum phenomena are crucial keys in obtaining the lower bound $(\eta/s)_{\mathrm{l.b.}}$ in Eq. (\ref{eq.LB}): These commonly bring about long lifetimes of Fermi and (molecular) Bose quasi-particles in the BCS and BEC regime, respectively. Then, $\eta$ is enhanced at low temperatures, bringing about a dip in the temperature dependence of $\eta/s$.
\par
\begin{figure}[t]
\centering
\includegraphics[width=6cm]{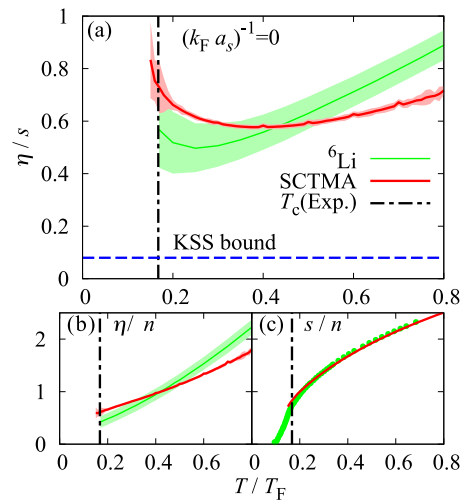}
\caption{(Color online) (a) Comparison of calculated $\eta/s$ ($T\ge T_{\mathrm{c}}$) in SCTMA with the recent experiment on a {\it mass-balanced} $^6$Li-$^6$Li unitary Fermi gas\cite{Bluhm:2017aa}. The latter result is obtained from the experimental data in Refs.\cite{Joseph:2015aa,Ku:2012aa}. $T_{\mathrm{c}}({\mathrm{Exp.}})$ is the observed $T_{\mathrm{c}}$\cite{Ku:2012aa}. (b) and (c) compare $\eta$ and $s$ in SCTMA with the experimental results\cite{Bluhm:2017aa,Ku:2012aa}, respectively.}
\label{fig8}
\end{figure}
\par
\begin{figure}[t]
\centering
\includegraphics[width=6cm]{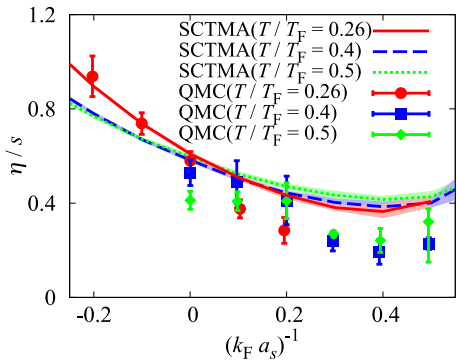}
\caption{(Color online) Comparison of our SCTMA results (lines) with the recent Quantum Monte Carlo results (points with error bars)\cite{Wlazlowski:2015aa}.
}
\label{fig9}
\end{figure}
\par
\subsection{Comparison with experiments}
\par
We finally compare our results with recent experiments on a mass-balanced $^{6}$Li-$^{6}$Li unitary Fermi gas\cite{Schafer:2009ab,Bluhm:2017aa}: The minimum value $(\eta/s)_{\mathrm{min}}^{\mathrm{unitarity}}=7.1$ (in the unit of $\hbar/(4\pi k_{\mathrm{B}})$) obtained in our SCTMA at the unitarity is somehow larger than $(\eta/s)_{\mathrm{min}}^{\mathrm{unitarity}}=6.3$ evaluated in Ref.\cite{Schafer:2009ab}, as well as $(\eta/s)_{\mathrm{min}}^{\mathrm{unitarity}}=5.0$ in Ref.\cite{Bluhm:2017aa}. As shown in Fig. \ref{fig8}(a), the calculated $\eta/s$ in SCTMA is comparable to the latter analysis\cite{Bluhm:2017aa}; however, the dip temperature $T_{\mathrm{min}}/T_{\mathrm{F}}\simeq 0.4$ in SCTMA is higher than the observed one $T_{\mathrm{min}}/T_{\mathrm{F}}\simeq 0.3$. We find from Figs. \ref{fig8}(b) and (c) that this discrepancy comes from, not $s$, but the detailed temperature dependence of $\eta$. 
\par
At present, $\eta/s$ around $(k_{\mathrm{F}}a_s)^{-1}=0.4$ has not been measured yet. However, it has recently been observed in a mass-balanced $^{6}$Li-$^6$Li Fermi gas that the minimum of $\eta$ exists in the BEC side ($0.25\lesssim (k_{\mathrm{F}}a_s)^{-1}\lesssim 0.5$)\cite{Elliott:2014ab}, which is consistent with our prediction. In the mass-balanced case, the deviation of the interaction strength at which $(\eta/s)_{\mathrm{l.b.}}$ is obtained from the unitary limit ($(k_{\mathrm{F}}a_s)^{-1}\simeq 0.4$) has also recently been obtained  by Quantum Monte Carlo simulation (QMC)\cite{Wlazlowski:2015aa} (see Fig. \ref{fig9}).
\par
\par
\section{Summary}
\par
To summarize, we have theoretically discussed the lower bound of the ratio $\eta/s$ and the effects of mass imbalance in the normal state of an ultracold Fermi gas in the BCS-BEC crossover region. For this purpose, we have calculated the shear viscosity $\eta$, as well as the entropy density $s$, within the same framework of the self-consistent $T$-matrix approximation, to numerically evaluate this ratio.
\par
We showed that the calculated $\eta/s$ does not contradict with the KSS conjecture: We obtained
\begin{align}
\frac{\eta}{s} \ge (4.5 \pm 0.2) \times \frac{\hbar}{4\pi k_{\mathrm{B}}}.
\label{eq.KSS}
\end{align}
The lowest value is obtained, not in the unitary limit, but slightly in the BEC regime at $(k_{\mathrm{F}}a_s)^{-1}\simeq 0.4>0$. Surprisingly, this lower bound is universal in a mass-imbalanced Fermi gas (within our numerical accuracy), in the sense that it is irrespective of the detailed value of $m_{\mathrm{L}}/m_{\mathrm{H}}$. We also pointed out that the Pauli blocking in the weak-coupling BCS regime, as well as the formation of two-body bound molecules in the strong-coupling BEC regime, are keys to obtaining this lower bound. 
\par
In a sense, the $m_{\mathrm{L}}/m_{\mathrm{H}}$-independence of the lower bound of $\eta/s$ is consistent with the prediction by KSS, although the value of the lower bound is about 4.5 times larger than their conjecture. So far, we have only numerically confirmed the universality of the lower bound of $\eta/s$ in a mass-imbalanced Fermi gas. If one can analytically prove it starting from the standard BCS model in Eq. (\ref{eq.3}), it would contribute to the further understanding of the KSS conjecture. Such an analytical approach remains as our future problem. In addition, because our results cover various kinds of Fermi atomic gases, ranging from mass-balanced $^6$Li-$^6$Li and $^{40}$K-$^{40}$K mixtures to a highly mass-imbalanced $^{40}$K-$^{161}$Dy mixture, systematic measurements of $\eta/s$ in these gases would also be an interesting experimental challenge. Although our numerical calculations in this paper cannot cover the more highly mass-imbalanced $^6$Li-$^{40}$K mixture, it would also be interesting to experimentally examine whether or not the same lower bound is obtained even in this case. Because the effects of mass imbalance are also important in Bose-Fermi mixtures and electron-hole systems, our results may contribute to the study of the KSS conjecture there.
\par
Although the present SCTMA approach still has room for improvement/extension, such as extension to the superfluid phase, as well as more sophisticated treatment of molecular correlations in the BEC regime, our results would be useful for, not only cold atom physics, but also various fields, such as condensed matter physics and high-energy QGP physics, where the KSS bound has extensively been discussed.
\par
\par
\section*{acknowledgments}
We thank D. Inotani for discussions. We also thank J. E. Thomas for informing us of Ref.\cite{Bluhm:2017aa}. This work was supported by KiPAS project at Keio University. D.K. was supported by KLL Ph. D. Program Research Grant from Keio University. Y.O. was supported by a Grant-in-aid for Scientific Research from MEXT and JSPS in Japan (No.JP18K11345, No.JP18H05406, and No.JP19K03689).
\par
\par


\begin{thebibliography}{10}
\bibitem{note} In the BCS regime, although $\eta$ again decreases near $T_{\mathrm{c}}$ (see Figs. \ref{fig2}(b1)-(b3)), $\eta/s$ is still very large because of the small values of $s$ there.
\end{thebibliography}
\end{document}